\documentclass[12pt]{article}
\usepackage{amsmath,amssymb}
\usepackage{graphicx,color}
\numberwithin{equation}{section}
\usepackage{cite}
\usepackage{bm}
\usepackage{dcolumn}
\newcommand{\be}{\begin{equation}}
\newcommand{\bea}{\begin{eqnarray}}
\newcommand{\eea}{\end{eqnarray}}
\newcommand{\ba}{\begin{array}}
\newcommand{\ea}{\end{array}}
\newcommand{\ee}{\end{equation}}
\newcommand{\nn}{\nonumber}

\expandafter\ifx\csname mathbbm\endcsname\relax

\else

\fi \textheight 22cm \textwidth 15cm \topmargin 1mm
\def\ft#1#2{{\textstyle{\frac{\scriptstyle #1}{\scriptstyle #2} } }}
\def\fft#1#2{{\frac{#1}{#2}}}

\def\0{{\sst{(0)}}}
\def\1{{\sst{(1)}}}
\def\2{{\sst{(2)}}}
\def\3{{\sst{(3)}}}
\def\4{{\sst{(4)}}}
\def\5{{\sst{(5)}}}
\def\6{{\sst{(6)}}}
\def\7{{\sst{(7)}}}
\def\8{{\sst{(8)}}}
\def\sst#1{{\scriptscriptstyle #1}}

\def\cG{{{\cal G}}}
\begin{document}
\begin{titlepage}
\vspace*{1mm}%
\hfill%
\vbox{
    \halign{#\hfil        \cr
                     } 
      }  
\vspace*{15mm}%
\begin{center}

{{\Large {\bf D-Dimensional Log Gravity  }}}

\vspace*{15mm} \vspace*{1mm} {Mohsen Alishahiha and Reza Fareghbal}

 \vspace*{1cm}

{\it  School of physics, Institute for Research in Fundamental Sciences (IPM)\\
P.O. Box 19395-5531, Tehran, Iran \\ }

\vspace*{.4cm}

email:  alishah@ipm.ir, fareghbal@theory.ipm.ac.ir

\vspace*{2cm}

\end{center}

\begin{abstract}
 We study Einstein gravity in dimensions $D\geq 4$ modified by curvature squared at critical point where
the theory contains only massless gravitons. We show that at the critical point a new mode
appears leading to a logarithmic gravity in the theory.  The corresponding logarithmic
solution may provide  a gravity description of logarithmic CFT in higher dimensions.
We note also that for special values of the parameters of the theory,  the model admits solutions
 with non-relativistic isometries.
\end{abstract}

\end{titlepage}

\section{Introduction}
Four dimensional  Einstein gravity  modified by curvature squared terms
 may provide a toy model to study quantum gravity in
four dimensions. The corresponding action of the theory is given
by\cite{{Stelle:1976gc},{Stelle:1977ry}}
 \be I = \fft{1}{2\kappa^2}\,
  \int \sqrt{-g}\,d^4x(R +\frac{6}{\ell^2} + \alpha  R^2+  \beta R^{\mu\nu} R_{\mu\nu} ) \,.\label{action}
\ee
When the cosmological constant is zero,  $\Lambda=0$, the theory
is renormalizable and  contains
massless  gravitons, massive spin 2  and a massive scalar fields\cite{{Stelle:1976gc},{Stelle:1977ry}}.  Nevertheless the theory has ghosts due to negative energy excitations
of the massive tensor. We note, however, that  at the special values of the parameters of the
theory either the massive tensor or the scalar would be absent. We note also that  when $\beta=0$ the model is unitary but non-renormalizable, while when the
curvature squared terms is given by the  Weyl tensor squared  the
model is neither unitary nor renormalizable.

With non-zero cosmological constant the model exhibits  new
features. Although in this case the theory still  contains
massless  gravitons, massive spin 2  and a massive scalar fields, there is a possibility
to tune the parameters such that only massless gravitons remain in the spectrum.

More precisely it has been shown \cite{Lu:2011zk}  that
for the special values of $\alpha$ and $\beta$ (critical point)   given by
 \be\label{cp}
  \beta=-3\alpha=-\frac{\ell^2}{ 2},
\ee
 the spin 2 field becomes massless and
at the same time the massive scalar is removed from the spectrum. As a result we are left
with four dimensional gravity with only massless gravitons.

In this letter we would like to further study the model at the
critical point. In particular we study  AdS wave solutions in
the mode and we will show that at the critical point it admits
logarithmic solutions. These solutions may provide gravity
descriptions for logarithmic CFT's in higher dimensions.

The letter is organized as follows. In the next section we will
consider AdS wave solutions in four dimensional theory at critical point. Generalization
to higher dimensions is presented in section three. The last section
is devoted to conclusions.

\section{AdS wave solution and Log gravity}

In this section we will study AdS wave solutions in the model given by the action \eqref{action}.
To procced we start with the equations of motion of the action \eqref{action}
 \be\label{eom}
  \cG_{\mu\nu} + E_{\mu\nu}=0\,,
\ee
where \cite{Lu:2011zk}
\bea
\cG_{\mu\nu} &=& R_{\mu\nu} -\ft12 R\, g_{\mu\nu} -\frac{ 3}{\ell^2}g_{\mu\nu}\,,
  \label{cGdef}\cr &&\cr
E_{\mu\nu} &=& 2\beta (R_{\mu\rho}\, R_\nu{}^\rho -\ft14 R^{\rho\sigma}
  R_{\rho\sigma}\, g_{\mu\nu}) + 2\alpha R\, (R_{\mu\nu} -\ft14 R\, g_{\mu\nu})
\cr &&\cr
&+&\beta\, ( \square R_{\mu\nu} + \ft12 \square R\,
  g_{\mu\nu} - 2\nabla_\rho \nabla_{(\mu} R_{\nu)}{}^\rho) +
 2\alpha\, (g_{\mu\nu}\, \square R -\nabla_\mu\nabla_\nu R)\,.\nn\label{Edef}
\eea
Since $E_{\mu\nu}$ vanishes for any Einstein space in four dimensions, with a negative
cosmological constant (as we choose here)
the model admits an AdS$_4$ vacuum solution whose radius is given by $\ell^2$.
This solution has been  studied in\cite{Lu:2011zk}.

Having  AdS vacuum solution, it is interesting to study AdS wave solutions
in the model\footnote{ AdS wave solutions for TMG and NMG models have been studied in
\cite{{AyonBeato:2004fq},{AyonBeato:2005qq},{AyonBeato:2009yq}}
 where it was shown that at the critical value of the parameters the solution develops
logarithmic behaviors. The same situation has also been observed in bi-gravity
and  Born-Infeld gravity \cite{{Afshar:2009rg},{Alishahiha:2010iq}}.}.
 To proceed we consider an anstaz for  AdS wave solutions as follows
\be
g_{\mu\nu}=\bar{g}_{\mu\nu}+F k_\mu k_\nu
\ee
where $k_\mu$ is a null vector field with respect to the metric $\bar{g}_{\mu\nu}$  with
$\bar{g}_{\mu\nu}$  being  the AdS$_4$ metric parametrized as
\be
ds^2=\frac{\ell^2}{r^2}\left(-2dx^+dx^-+dy^2+dr^2\right).
\ee
Note that $F$ is an  arbitrary function which is independent of the integral parameter along
$k_\mu$.  In other words the ansatz may be given by
\be\label{ansatz}
ds^2=\frac{\ell^2}{r^2}\left(-F(x^+,y,r)d{x^+}^2-2dx^+dx^-+dy^2+dr^2\right).
\ee
Plugging this ansatz into the equations of motion \eqref{eom} one finds
\bea
\bigg[\frac{\beta r^3}{\ell^2} \left(\frac{\partial^4}{\partial r^4}
+2\frac{\partial^2}{\partial y^2}\frac{\partial^2}{\partial r^2}
+\frac{\partial^4}{\partial y^4}\right)&+&\left(1-\frac{8(3\alpha+\beta)}{\ell^2}\right)\\
&\times& \left(r\frac{\partial^2}{\partial
r^2}-2\frac{\partial}{\partial r}+r\frac{\partial^2} {\partial
y^2}\right)\bigg]F(x^+,r,y)=0.\nn \eea
For simplicity we assume that
$F$ is independent of $y$ coordinate. In this case the  above
equation reads
\be \left[\frac{\beta r^3}{\ell^2}
\frac{\partial^4}{\partial
r^4}+\left(1-\frac{8(3\alpha+\beta)}{\ell^2}\right)\left(r\frac{\partial^2}{\partial
r^2}-2\frac{\partial}{\partial r} \right)\right]F(x^+,r)=0.
\ee
It is clear that the most general solution of the above differential
equation is in the form of $r^x$ with constant $x$ satisfying the
following characteristic equation
\be\label{ch} x(x-3)\left[\beta
(x-1)(x-2)-8(3\alpha+\beta)+\ell^2\right]=0.
\ee
Therefore a generic solution of the equations of motion is\footnote{A $y$ dependent solution can also be obtained as
$F(x^+,y,r)=(c_0(x^+)+c_1(x^+) y)F(x^+,r)$.}
 \be
F(x^+,r)=f_4(x^+)+f_3(x^+)r^3+f_2(x^+)r^{\Delta_+}
+f_1(x^+)r^{\Delta_-}.
\ee
where
$\Delta_\pm=\frac{3}{2}\pm\sqrt{\frac{9}{4}-\frac{\ell^2-24\alpha-6\beta}{\beta}}$, and
$f_i$'s are undetermined functions of $x^+$.

It is natural to look for a possibility of having multiplicities in
the roots of the characteristic equation. Actually we observe that
at the critical point where the parameters $\alpha$ and $\beta$ are
given by the equation \eqref{cp}  the characteristic equation
degenerates leading to  new logarithmic solutions as follows
\be
F(x^+,r)=f_4(x^+)+f_3(x^+)r^3+\left[f_2(x^+)+f_1(x^+)r^3\right]\log(r).
\ee
In other words at the critical point the model admits a new vacuum
solution which is not asymptotically locally AdS$_4$. Therefore in
order  to accommodate the new solution one needs to change the
asymptotic behavior of the metric. More precisely using the
Fefferman-Graham coordinates for the metric
\be
ds^2=\frac{dr^2}{r^2}+\frac{1}{r^2}g_{ij}(x_i)dx^idx^j,
\ee the
equations of motion give a possibility to have a wider class of
boundary conditions for the metric as follows
\be
g_{ij}=b_{(0)\;ij}\log(r)+g_{(0)\;ij}+g_{(3)\;ij}r^3+b_{(3)\;ij}r^3\log(r)+\cdots\;.
\ee
 Typically when $b_{(0)\;ij}$ is non zero, where the
solution is not an  asymptotically locally AdS$_4$,  to maintain the
variational principle well posed with the Dirichlet boundary
condition one needs to modify the variational principle by imposing
an additional boundary condition\cite{Skenderis:2009nt}.
Indeed from the above expression
for the asymptotic behavior of the metric one finds
\be
g_{(0)\;ij}=\lim_{r\rightarrow
0}\left(g_{ij}-r\log(r)\;\frac{\partial g_{ij}}{\partial r}\right),
\;\;\;\;\;\;\;\;\;\;b_{(0)\;ij}=\lim_{r\rightarrow 0}r\frac{\partial
g_{ij}}{\partial r},
\ee
showing that in order to fix the boundary
conditions not only one needs the value of the boundary metric but
also  its radial derivative.

In the context of  AdS/CFT correspondence \cite{Maldacena:1997re}
both $g_{(0)\;ij}$ and $b_{(0)\;ij}$ may be treated as two  sources
for two operators in the boundary 3-dimensional CFT. We note,
however, that since in the presence of non-zero $b_{(0)\;ij}$ the
geometry is not asymptotically locally AdS$_4$, the parameter
$b_{(0)\;ij}$ should be considered as a source for an irrelevant
operator in the dual CFT (see for example \cite{Skenderis:2009nt}).
Nevertheless for a sufficiently small $b_{(0)\;ij}$ one  could still
use the AdS/CFT correspondence to describe the dual theory which is
expected to be a logarithmic CFT\footnote{Logarithmic CFT's in the context of AdS/CFT
correspondence have also been studied in\cite{Ghezelbash:1998}. We note, however, that in
this paper the  authors have fixed the background to be $AdS$ and the logarithmic
behavior comes for the specific action of the fields in the bulk.}.

 As a result the logarithmic
solution of the action \eqref{action}  indicates that critical
gravity gives  a  gravity
description for three dimensional logarithmic CFT's. Actually the
situation is similar to those in TMG and NMG models where it is
believed  that the dual theory is  logarithmic CFT
\cite{{Grumiller:2008qz},{Grumiller:2009sn}} at critical point.  Following
\cite{{Skenderis:2009nt},{Alishahiha:2010bw}} it would be interesting to find  two point
functions and the corresponding  new anomaly parameter in this
model using holographic renormalization method.

As a final remark  we note that away from the critical point and
for a specific values of $\alpha$ and $\beta$, the model admits another  one parameter
solution. Indeed when
\be
24\alpha=2(2n^2-n-4)\beta+\ell^2,
\ee
with $n$ being a free parameter, for a specific choice of the intergral constants one finds the
following solution
\be
ds^2=\frac{\ell^2}{r^2}\left(-\frac{dt^2}{r^{2n-2}}-2dtd\xi+dy^2+dr^2\right).
\ee
We recognize this solution as a gravity solution whose dual theory is a non-relativistic
field theory \cite{Son:2008ye}\footnote{Lifshitz black holes in Einstein gravity with
curvature squared terms have also been studied in\cite{{Cai:2009ac},
{AyonBeato:2010tm}}.} .
 In particular for $n=2$ the isometry of the metric is
Schr\"odinger group and the solution provide a gravity description
for a non-relativistic CFT.

It is worth mentioning that in our model the non-relativistic
holographic solution is obtained in a pure gravitational system,
though the one studied in \cite{Son:2008ye}  has been obtained in a
model which contains a gravity coupled to  a massive gauge field. Of
course we have not checked whether this vacuum is stable in the
sense that  small fluctuations above it  have  non-negative mass spectrum.

\section{Higher Dimensions}

In this section we would like to extend our previous discussions to
higher dimensions. Recently D-dimensional extended gravities have
been studied in  \cite{Deser:2011xc} where the authors have
considered the following gravitational action
\begin{equation}
I=\frac{1}{\kappa}\int d^{D}x\,\sqrt{-g}\left[R-2\Lambda_{0}+\alpha R^{2}+\beta R^{\mu\nu}R_{\mu\nu}+\gamma\left(R^{\mu\nu\rho\sigma}R_{\mu\nu\rho\sigma}-4R^{\mu\nu}R_{\mu\nu}
+R^{2}\right)\right].\label{eq:Quadratic_action}
\end{equation}
The corresponding equations of motion are \cite{Gullu:2009vy}
\begin{eqnarray}
&&R_{\mu\nu}-\frac{1}{2}g_{\mu\nu}R + \Lambda_0 g_{\mu\nu}
+2\alpha R\left(R_{\mu\nu} -\frac{1}{4}g_{\mu\nu}R\right)+\left(2\alpha+\beta\right)\left(g_{\mu\nu}\square-\nabla_{\mu}\nabla_{\nu}\right)R\nonumber\\&&
+2\gamma\left[RR_{\mu\nu}-2R_{\mu\sigma\nu\rho}R^{\sigma\rho}+R_{\mu\sigma\rho\tau}R_{\nu}^{\;\;\sigma\rho\tau}
-2R_{\mu\sigma}R_{\nu}^{\;\;\sigma}-\frac{1}{4}g_{\mu\nu}\left(R_{\tau\lambda\sigma\rho}^{2}-4R_{\sigma\rho}^{2}
+R^{2}\right)\right]\nonumber\\&&+\beta\square\left(R_{\mu\nu}-
\frac{1}{2}g_{\mu\nu}R\right)+2\beta\left(R_{\mu\sigma\nu\rho}
-\frac{1}{4}g_{\mu\nu}R_{\sigma\rho}\right)R^{\sigma\rho} =0 .
\label{fieldequations}
\end{eqnarray}
For generic values of the parameters $\Lambda, \alpha,\beta$ and
$\gamma$ the model has two distinct vacua such that
$R_{\mu\nu}=\frac{2\Lambda}{D-2}g_{\mu\nu}$, where $\Lambda$ is a
root of the following equation\cite{Deser:2011xc}
\be\label{Lam}
\Lambda_0-\Lambda=2\Lambda^2\big[(D\alpha+\beta)\frac{D-4}{(D-2)^2}+
\frac{(D-3)(D-4)}{(D-1)(D-2)}\gamma\big].
\ee
It is always possible
to tune the parameters such that at least one of the vacua to be an
AdS$_D$  geometry. In this case the radius of the AdS geometry is
given in terms of $\Lambda$ as follows\footnote{When the right hand
side of the equation \eqref{Lam} vanishes
with the assumption of negative cosmological constant, $\Lambda=\Lambda_0<0$, the model
admits a unique AdS solution. In this case when $D\neq 4$ the
parameters of the model have to obey the constraint
$D\alpha+\beta+\frac{(D-2)(D-3)}{D-1}\gamma=0$.}
\be
\ell^2=-\frac{(D-1)(D-2)}{2\Lambda}.
\ee
It was shown in
\cite{Deser:2011xc} that for appropriate choice of the parameters
there exists a critical point where \be\label{cpD}
\beta=-{4(D-1)\over D}\,\alpha,\;\;\;\;
\frac{(D-1)(D-2)}{4(-\Lambda)}=
(D-1)(D\alpha+\beta)+(D-3)(D-4)\gamma,
\ee
 at which the model has only massless tensor gravitons.

Following our discussions in the previous section it is natural to
look for  AdS wave solutions in the model and in particular to see
if the model supports a logarithmic solution at the critical point.
To proceed we consider an ansatz as follows
 \be\label{metric higher}
ds^2=\frac{\ell^2}{
r^2}\left(-F(x^+,r,x_i)du^2-2dx^+dx^-+dr^2+(dx_i)^2\right). \ee
Plugging this ansatz into the equations of motion one finds
\bea\label{F equation higher dimension} &&\Bigg\{\frac{\beta
r^3}{D-2}\left(\frac{\partial^2}{\partial
r^2}+\frac{\partial^2}{\partial
x_i^2}\right)^2-\frac{2(D-4)}{D-2}\beta r^2
\left(\frac{\partial^2}{\partial r^2}+\frac{\partial^2}{\partial
x_i^2}\right) \frac{\partial}{\partial r} +\beta
r(D-4)\frac{\partial^2}{\partial r^2}\\ \nonumber  &&\quad +\frac{
r}{D-2}\left[\ell^2-2D(D-1)\alpha-4(D-2)\beta-2(D-3)(D-4)\gamma\right]\left(\frac{\partial^2}{\partial
r^2}+\frac{\partial^2}{\partial x_i^2}\right)\\ \nonumber && \quad
-\left[\ell^2-2D(D-1)\alpha-(3D-4)\beta-2(D-4)(D-3)\gamma\right]
\frac{\partial}{\partial r}\Bigg\}F(x^+,r,x_i)=0. \eea The equation
may be simplified with the assumption that the function $F$ is
independent of transverse directions $x_i$'s. In this case  a
generic solution of the resultant equation will be of the form $r^x$
for constant $x$. From the above equation the  characteristic
equation reads \be \frac{\beta}{D-2}x(x-D+1)\left( x^2-
(D-1)x+\frac{A}{\beta}\right)=0 \ee where
$A=l^2-2(D\alpha+\beta)(D-1)-2(D-3)(D-4)\gamma$.

Therefore the most general solution of the equations of motion is
\be
F(x^+,r)=f_4(x^+)+f_3(x^+)r^{D-1}+f_2(x^+)r^{\Delta_+}
+f_1(x^+)r^{\Delta_-},
\ee
where $\Delta_\pm=\frac{D-1}{2}\pm\sqrt{(\frac{D-1}{2})^2-\frac{A}{\beta}}$, and
$f_i$'s are undetermined functions of $x^+$.

 We note that at the critical point
\eqref{cpD} where $A=0$ the characteristic equation degenerates leading to new
logarithmic solutions as follows
\be
F(x^+,r)=f_4(x^+)+f_3(x^+)r^{D-1}+\left[f_2(x^+)
+f_1(x^+)r^{D-1}\right]\log(r).
\ee
As a results, following our
discussions in the previous section, the gravitational model based
on the action \eqref{eq:Quadratic_action} at the critical point
 may provide a gravity description for $D-1$
dimensional logarithmic CFT's.

The model also admits non-relativistic solutions when the parameters of the model obey
the constraint $A+2(n-1)(2n-D-1)\beta=0$ for $n\neq1$. In this case
for special choice of the integral constnats one finds
\be
ds^2=\frac{\ell^2}{r^2}\left(-\frac{dt^2}{r^{2n-2}}-2dtd\xi+dr^2+(dx_i)^2\right).
\ee
\section{Conclusions}
In this letter we have studied AdS wave solutions in $D$-dimensional
Einstein gravity with curvature squared modification.
At the critical point where the theory contains only
massless gravitons the model admits logarithmic solutions.

 We have also shown that for special values of the parameters of the model,
one could  have non-relativistic solutions. In particular the model admits solutions with Schr\"odinger isometry. Therefore these models could provide gravity descriptions for non-relativistic and logarithmic CFT's.

It is important to note that the existence of  these solutions are due to the curvature
squared terms in the action.  Actually restricting to four dimensions we note that any Einstein solutions are the solutions of the model with curvature squared action.  In fact black
 hole solutions of the Eisntein kind have altready been discussed in \cite{Lu:2011zk} where
 the authors have shown that the corresponding black holes have zero mass and entropy!

 It would be interesting to find other
solutions in the model which  are not Einstein solutions. In particular one may seek for
non-Einstein  black hole solutions in the model. These black holes could be of the
logarithmic solution as well.

\vspace{.5cm}

{\bf Note added}:
After we submitted our paper to arXiv, two other papers,
\cite{Gullu:2011sj} and \cite{Bergshoeff:2011ri}, appeared in arXiv where the
similar logarithmic solutions have been discussed. It was also conjectured that
the corresponding dual field theory could be a logarithmic CFT.

\section*{Acknowledgments}

We would like to thank Ali Naseh for discussions. This work is supported by Iran National Science Foundation (INSF). We would also like to thank referee for his/her useful comments.

\end{document}